\documentclass[useAMS,usenatbib]{mn2e}
\usepackage{epsfig}
\usepackage{aas_macros}

\def\lesssim{\lower.5ex\hbox{$\; \buildrel < \over \sim \;$}}
\def\gtrsim{\lower.5ex\hbox{$\; \buildrel > \over \sim \;$}}

\title[Modeling gamma-ray burst observations]{Modeling gamma-ray burst observations by Fermi and MAGIC including attenuation due to diffuse background light}

\author[R. C. Gilmore et al.]{Rudy C. Gilmore$^{1}$\thanks{E-mail:rgilmore@physics.ucsc.edu}, Francisco Prada$^{2,3}$, Joel Primack$^{1}$\\
$^{1}$Department of Physics, University of California, Santa Cruz, CA 95064 \\
$^{2}$Instituto de Astrof\'isica de Andaluc\'ia, CSIC, Apdo. Correos 3004, E-18080 Granada, Spain \\
$^{3}$Visiting Research Physicist at the Santa Cruz Institute for Particle Physics (SCIPP), University of California, Santa Cruz, CA 95064}

\begin{document}
\date{\today}

\def\lesssim{\lower.5ex\hbox{$\; \buildrel < \over \sim \;$}}

\pagerange{\pageref{firstpage}--\pageref{lastpage}} \pubyear{2009}

\maketitle
\label{firstpage}

\begin{abstract}
Gamma rays from extragalactic sources are attenuated by pair-production interactions with diffuse photons of the extragalactic background light (EBL).  Gamma-ray bursts (GRBs) are a source of high-redshift photons above 10 GeV, and could be therefore useful as a probe of the evolving UV background radiation.  In this paper, we develop a simple phenomenological model for the number and redshift distribution of gamma-ray bursts that can be seen at GeV energies with the {\it Fermi} satellite and MAGIC atmospheric Cherenkov telescope.  We estimate the observed number of gamma rays per year, and show how this result is modified by considering interactions with different realizations of the evolving EBL.  We also discuss the bright {\it Fermi} GRB 080916C in the context of this model.  We find that the LAT on {\it Fermi} can be expected to see a small number of photons above 10 GeV each year from distant GRBs.  Annual results for ground-based instruments like MAGIC are highly variable due to the low duty cycle and sky coverage of the telescope.  However, successfully viewing a bright or intermediate GRB from the ground could provide hundreds of photons from high redshift, which would almost certainly be extremely useful in constraining both GRB physics and the high-redshift EBL. 

\end{abstract}


%\label{grbatt}

\section{Introduction}
High-energy gamma rays traveling cosmological distances can interact with the diffuse UV, optical, and IR radiation fields in electron-positron pair production interactions, leading to an effective optical depth for VHE sources \citep{nikishov62,gould&schreder67}.  A recent technique in gamma-ray astronomy has been to use observations of relatively nearby blazars observed by Imaging Atmospheric Cherenkov Telescopes (IACTs) to constrain the EBL (e.g. \citealp{aharonian06,mazin&raue07,albert08}).

While blazars have been the primary target of efforts to detect the effects of EBL attenuation in high-energy spectra, another exciting possibility is to see these same effects in observations of GRBs.  Until recently, limitations on the effective areas and energy ranges of high-energy experiments, and their ability to respond sufficiently quickly to transient events have hindered observations at GeV energies that might reveal evidence of this phenomenon.

The EGRET experiment on the {\it Compton Gamma-ray Observatory} ({\it CGRO}) operated with energy range 20 MeV--30 GeV and effective area $\sim$1000 cm$^2$.  This experiment detected a total of 5 bursts above 30 MeV in 4 years of operation, including 4 individual photons above 1 GeV \citep{dingus95}.  While these EGRET detections do suggest the presence of a very-high energy component in the spectrum of some GRBs, it is difficult to draw more conclusions due to the small number of high-energy gamma rays detected.  At the same time, the BATSE instrument on CGRO detected thousands of GRBs at energies between 20 keV and 2 MeV \citep{paciesas99}.  The {\it Swift} mission has been finding bursts at a rate of about 8 per month since its launch in December 2004 at energies between 15 and 150 keV \citep{sakamoto08}.   

Until the launch of the {\it Fermi Gamma-ray Space Telescope} in 2008, the only other possibility to view high-energy emission from GRBs was with ground-based experiments -- IACTs such as H.E.S.S., MAGIC, or VERITAS, or air shower arrays such as Milagro.  While IACTs have the advantage of much larger effective collecting areas than satellite detectors, they are limited by their low duty cycles and small fields of view, which make a serendipitous GRB detection very unlikely.  Therefore, these IACTs must be alerted to a burst event by another detector, usually a gamma- or x-ray satellite, and slew to its position.  This introduces a new technical limitation on ground-based observations; the delay time in receiving an alert and moving the telescope means that much or all of the primary emission of a burst can be missed.  Air shower arrays like Milagro do not have slewing issues, but are generally not as sensitive as their IACT counterparts, particularly at lower energies. 

Despite the difficulties involved, followup observations have been made of many GRBs by all major experiments mentioned above.  MAGIC responded to 35 burst alerts between January 2005 and June 2008, a rate of about 1 per month, with an average slew time of 45 seconds (\citealp{garczarczyk08}, see also \citealp{albert07d}).  These attempts were only able to place upper limits on the flux.   Negative results were also found in 32 observations over 4 years by H.E.S.S. \citep{aharonian09}.  Most of these observations did not begin until several hours after the initial detection of the event.  VERITAS has reported limits for a small number of bursts \citep{horan08}.  Air shower arrays are generally less sensitive than IACTs, but have the advantage of much larger fields of view and duty cycles.  The Milagro prototype, Milagrito, claimed a possible detection of prompt emission from GRB 970417A at $>650$ GeV energies \citep{atkins03}.  The redshift of this burst is not known, but to be detected at these energies it would have had to have been quite nearby, as the universe becomes optically thick due to EBL attenuation at low redshift ($z<0.2$ for our lowest EBL model) for $E_\gamma > 650$ GeV.

In this work, we focus on two telescopes that are well-suited to detect GeV emission from GRBs: {\it Fermi} and the Major Atmospheric Gamma-ray Imaging Cherenkov Telescope (MAGIC).  {\it Fermi} contains two instruments.  The GLAST Burst Monitor (GBM) is designed for finding prompt emission from GRB from 10 keV to 30 MeV, and the Large Area Telescope (LAT) views gamma rays from 20 MeV to 300 GeV with an effective area peaking at $\sim 9000$ cm$^2$, giving it both detection area and energy maximum about 10 times those of EGRET.  {\it Fermi} has been operated in survey mode, in which it views the entire sky every 3 hours, since shortly after its launch, and will continue to be in this mode for most of its $>$5-year lifespan.  The MAGIC experiment was recently upgraded to its second phase, MAGIC-II, and now consists of 2 large IACTs, each with a mirror area 236 m$^2$ (the effective detection area for gamma rays is much larger, $> 10^5$ m$^2$ at optimal energies).  The MAGIC telescopes are designed to have a low energy trigger threshold, $<$50 GeV near zenith, and are capable of repositioning to any point on the sky within 30 seconds \citep{bastieri05,albert07d}.  In this work, we will model the observational capabilities of stereoscopic observations with this telescope using two different assumptions about instrument sensitivity at low energies.

In Section \ref{model}, we present a model for high energy emission from GRBs based on {\it Swift} and {\it CGRO} observations.  We also briefly review recent predictions for the cosmological gamma-ray opacity due to pair-production interactions with UV and optical background photons, and the properties of the instruments for which we are predicting detection rates.  Our main results are presented in Section \ref{results}, where we show annual predictions for high-energy gamma-ray counts from GRBs for {\it Fermi} (Section \ref{fermipred}) and MAGIC (Section \ref{magicpred}).  In Section \ref{disc}, the impact of our findings for future observations and EBL constraints is summarized.  We also review our model in the context of the bright {\it Fermi} GRB 080916C, and calculate the number of counts that could have been seen by the MAGIC telescope had this burst fallen within its field of view. 

\section{Model}
\label{model}

In order to estimate the number of GeV gamma-rays from GRBs that will be available to {\it Fermi} and MAGIC, we develop a simple model to estimate the fluence that could be seen by these experiments over a given time.  As only bursts with known redshift are useful to our ultimate goal of probing UV and optical background fields via attenuation effects, we base our analysis on the population of bursts observed by the {\it Swift} Burst Alert Telescope (BAT) with measured redshift.  Data for these bursts has been taken from the {\it Swift} GRB table available online \footnote{http://heasarc.gsfc.nasa.gov/docs/swift/archive/grb\_table/}.  Our model rests upon two assumptions about high-energy emission: {\bf (1)} that the population of bursts seen by {\it Fermi} and MAGIC with redshifts that are eventually determined will be similar to these {\it Swift} bursts in number and fluence statistics, and {\bf (2)} that these bursts produce high energy emission, both prompt and afterglow, with a fluence that is proportional to that observed in the BAT energy range, 15--150 keV.  Here we define prompt emission as being approximately synchronous with the bulk of the flux seen by the BAT ($T_{90}$, as discussed below), and the afterglow as occuring in the minutes to hours afterward.  Our model is purely observational and phenomenological, and does not attempt to quantify in any way the intrinsic parameters of the bursts, nor do we make assumptions about the actual population statistics of these events.

\subsection{GRB Emission}
\label{grbemiss}
The populations of bursts seen by BATSE and {\it Swift} were analyzed in \citet{dai09}, who argued that these populations were similar, and that there was no difference between the subsets of optically detected {\it Swift} bursts with and without redshifts.  This suggests that the events for which we now have redshift information are not different from the GRB population as a whole.  We necessarily exclude from our analysis the population of `dark bursts' \citep*{virgili09}, for which it is not possible to measure redshifts due to lack of an optical afterglow.  We note that there have been suggestions that the low luminosity population of bursts are distinct from their brighter counterparts \citep{liang07}.  The BATSE sample should be very similar to that viewed by GBM on {\it Fermi}, and this should enable a test of assumption (1).

\begin{figure*}
\psfig{file=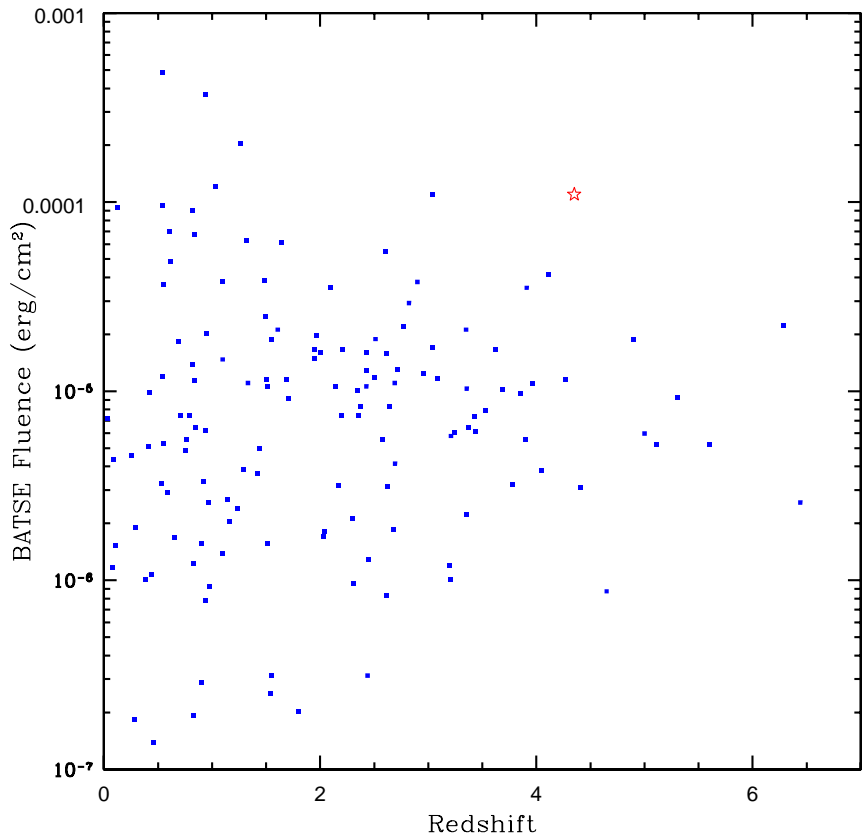,width=12cm}
\caption{The bursts seen by {\it Swift} BAT for which we have well-measured redshifts, shown on axes of redshift and fluence.  The fluences observed by BAT at energies from 15--150 keV have been converted to the BATSE energy range (20 keV -- 2 MeV) using a Band function over these energy ranges with peak 250 keV and low and high energy spectral indices of -1 and -2.2, respectively.  For reference, we also show the position of GRB 080916C on this plot based on its GBM fluence (single red star).  This burst was not seen by  {\it Swift} until nearly 17 hours after its detection by GBM and LAT \citep{kennea08}, and was not included in our analysis.}
\label{fig:batfluxdist}
\end{figure*}

We use the 145 bursts that were observed by {\it Swift} BAT between January 2005 and June 2009 and have known redshift.  Figure \ref{fig:batfluxdist} shows these bursts plotted as a function of redshift and BAT fluence.  A considerable amount of fluence from these bursts arises largely from a few bright events; the brightest 10 per cent of bursts in the sample accounts for approximately 55 per cent of the fluence.  While high energy flux has only been seen from a handful of bright GRBs using EGRET, the fact that these bursts account for a large fraction of fluence seen at lower energies means that our assumption (2) should be reasonable even if the proportionality does not hold for faint bursts.  We have not included {\it Fermi} LAT bursts such as GRB080916C in this analysis, although we do show where this event would have been in Fig.~\ref{fig:batfluxdist} based on its GBM fluence.  We will discuss this event in the context of our emission model in Section \ref{080916C}. 

Two recent papers have related the keV/MeV flux from GRBs to high energy emission, and have estimated the ratio of fluences in these regimes.  \citet{le&dermer09} estimated the count rate for GeV photons in the {\it Fermi} LAT based on the bursts seen by the EGRET spark chamber.  The $F_{LAT}/F_{BATSE}$ fluence ratio inferred there varied from 0.05 to over 0.3.  Based on the deadtime factors affecting some EGRET GRB observations these authors argue that a ratio of greater than 30 per cent between BATSE and EGRET is reasonable.  \citet*{ando08} made the assumption that there is a log-normal distribution of $F_{GeV}/F_{MeV} \approx F_{EGRET}/F_{BATSE}$ in the roughly 100 BATSE bursts that were in the field of view of EGRET.  A maximum likelihood fit to the available data suggested a ratio of $0.003 \leq F_{GeV}/F_{MeV} \leq 0.06$.  Slightly different assumptions about the high-energy spectrum and energy range of the GeV emission were used in each case.  Le \& Dermer used an index of -2, similar to the best fit to the EGRET bursts of -1.95 from \citet{dingus95}, while the Ando et al.~work assumed an spectral index of -2.4 in the EGRET energy range.  The latter noted that hardening the spectral index could have increased their values for the flux ratio coefficient significantly.    

In this paper we use the energy fluences seen by {\it Swift} BAT to predict GeV emission using the ratio 
\begin{equation}
\rho \equiv F_{\mbox{\scriptsize EGRET}}/F_{\mbox{\scriptsize BATSE}}
\end{equation}
where $F_{\mbox{\scriptsize EGRET}}$ and $F_{\mbox{\scriptsize BATSE}}$ refer to the fluence (time-integrated flux over the duration of the event) across the EGRET spark chamber and BATSE energy ranges, taken to be 100 MeV -- 5 GeV and 20 keV -- 2 MeV respectively.  A constant spectral index is assumed to be valid from the EGRET energy range up to $>100$ GeV.  We take an approach similar to Le \& Dermer and use a value of $\rho=0.1$ for prompt phase emission, which does a reasonable job matching the high energy fluence seen for the recent GRB 080916C, see Section \ref{080916C}.  This is higher than the range of values for $\rho$ proposed by Ando et al., but that paper also assumed a softer high energy spectrum in deriving results.  

Afterglows are also a possible source of high-energy emission, though one that is even more poorly constrained than the prompt phase.  There are various mechanisms that have been hypothesized as possible sources of GeV photons.  A popular assumption invokes inverse-Compton upscattering of synchrotron photons in the GRB outflow (SSC mechanism), although a variety of other sources are possible, such as SSC emission from the internal x-ray flares seen in afterglows or Compton upscattering of these photons by electrons accelerated in the external shock \citep{fan08}.  Limits from EGRET observations suggest a typical fluence ratio of 0.01 to 0.1, and a spectral index of -1.5 to -2 \citep{ando08}.  For afterglows, we assume a ratio of $\rho=0.01$ in this work.  

In converting the fluence seen by {\it Swift} BAT (15 -- 150 keV) to BATSE flux, we assume a common Band \citep{band93} functional form over the BAT--BATSE energy range:
\begin{eqnarray}
\lefteqn{\frac{dN}{dE} = A_0 \: [ E^{\alpha_1} \: e^{-\frac{E}{E_{br}}(\alpha_1-\alpha_2)} \: \Theta(E_{br}-E)} \nonumber \\
& & + E_{br}^{\alpha_1-\alpha_2} \: e^{\alpha_2-\alpha_1} \: E^{\alpha_2} \: \Theta(E-E_{br})].
\end{eqnarray}
\noindent Here $\alpha_1$ and $\alpha_2$ are the low and high energy indices, $E_{br}$ is the break energy, and $\Theta$ is the Heaviside step function.  Unfortunately, the relatively narrow energy band of the BAT does not allow one to resolve the structure of the Band peak in most cases for the {\it Swift} sample.  We use -1 and -2.2 for the low and high indices, and assume a break energy of 250 keV; these are the typical values seen in an analysis of BATSE bursts by \citet{preece00}, and lead to a ratio of 4.6 between the BATSE and BAT fluences.  The high energy flux in this model, effective in the EGRET energy range and at GeV energies, is taken to be a power law with a normalization found from the MeV--GeV proportionality, and is separate from the high-energy Band slope $\alpha_2$.  The GeV flux is then
\begin{equation}
\frac{dF}{dE}= \frac{\rho F_{\mbox{\scriptsize BATSE}}}{\int_{\mbox{\scriptsize EGRET}} E^{\beta} \: dE} \: E^{\beta} 
\label{eq:heflux}
\end{equation}
where $\beta$ is the high-energy spectral index and is determined independently of the Band function parameters.  We assume a high-energy spectral index of $-1.95$ for prompt phase photons, consistent with the EGRET results, and a harder spectral index of $-1.5$ for the afterglow component, which would occur if the spectral peak from inverse Compton emission is at energies equal to or higher than those being observed.   We do not discriminate between long and short bursts in our analysis.  As can be seen in Figure \ref{fig:slewfact}, only a small fraction of the bursts in our sample can be classified as short ($T_{90} \leq 2$ sec), and their contribution to the total flux is very low.

The inverse-Compton scattering of photons to high energies is limited by Klein-Nishina suppression, which reduces the flux of photons at energies that are higher than the electron rest mass $m_e$ in the particle's rest frame.  We will not include a possible cutoff due to this effect in our calculation, but will reserve discussions of the implications until Section \ref{intcut}.  In the simple SSC case \citep{chiang&dermer99,zhang&meszaros01,sari&esin01}, this can be written as 
\begin{equation}
E_{KN} \gtrsim \Gamma_b \gamma_e m_e c^2,
\end{equation}
where $\Gamma_b$ is the bulk Lorentz factor of the outflow, and $\gamma_e$ is the typical gamma factor for the electrons responsible for the synchrotron peak.  Including the effect of redshift, the observed gamma-ray energies affected by this suppression are \citep{panaitescu08}
\begin{equation}
\label{eq:kncut}
E_{obs} \gtrsim \frac{\Gamma_b \gamma_e}{1700 (1+z)} \mbox{ GeV}.
\end{equation}
\noindent Constraints from beaming and escape of high-energy radiation suggest large Lorentz factors for the bulk flow of material in the prompt and early afterglow phases of bursts, $\Gamma_b \sim 100$ \citep{meszaros06}.  The electrons will typically have a power law distribution in energy determined by the cooling rate and therefore Klein-Nishina effects do not lead to an abrupt spectral cutoff, but if the typical Lorentz factor is sufficiently low then observations in the 10--100 GeV energy range could be impacted.  As mentioned in \citet{ando08}, the electron Lorentz factor in external shocks in the afterglow is expected to be higher than that of the prompt emission, and therefore this may be a more likely source of detectable high energy photons.

\subsection{Cosmological gamma-ray opacity}
In this work we consider the effect of the UV-optical background on high-energy GRB observations, using three different models of the evolving background.  These models are based on the work of \citet{gilmoreUV}, in which semi-analytic models of galaxy formation and evolution are combined with estimates of quasar emissivity, ionizing escape fraction from star-forming galaxies, and processing of ionizing photons by the inter-galactic medium to accurately compute both ionizing and non-ionizing fluxes.  The low model in this work was based upon a WMAP3 cosmology that leads to delayed structure formation, due primarily to a low value for the dark matter power spectrum normalization, $\sigma_8 = 0.761$.  The star-formation history in this model peaks near redshift 2 and is in agreement with the lowest values determined for high-redshift star formation.  The rapid decline in star-formation rate density in this model leads to a UV background that is consistent with Ly$\alpha$ ionization measurements only with a high escape fraction of ionizing photons from galaxies, indicating that this is the minimal level of UV background which can be consistent with high redshift observations.  The fiducial model is based upon a semi-analytic model that uses a concordance cosmology ($\Omega_\Lambda = 0.7$, $\Omega_m =0.3$, $\sigma_8 = 0.9$).  This model produces a UV luminosity density in good agreement with measurements at low and high redshift, and a star-formation rate peaking at $z \approx 2.75$.  Finally, the fiducial high-peaked (`high') model features a star formation rate density and UV emissivity that is higher than that of the fiducial model at $z>3$.  The star-formation history in this model peaks at $z=5$, and is consistent with some of the higher determinations of star-formation rate density above redshift 3.  Because this model provides maximum stellar emission at high redshift, and produces a stellar mass density exceeding most observations, this should be considered a somewhat extreme model  resulting in maximal background levels.  This model converges to the fiducial model at redshifts lower than 3, and opacities for gamma-ray sources closer than this are not significantly different from the fiducial case.

\subsection{Instrument Properties}
\label{instprops}
For our purposes the performance of the LAT instrument on {\it Fermi} can be described with a relatively small number of parameters.  We take the LAT to have an effective area of 9000 cm$^2$ up to an energy of 300 GeV.  While this may be an overestimate of the effective area at the highest energies, the number of photons seen at these energies is likely to be negligible.  The integrated field of view for {\it Fermi} is found to be approximately $\sim 20500$ cm$^2$ sr; we therefore assume a field of view $20500/9000 \approx 2.28$ sr.  It is conservatively assumed in our analysis that {\it Fermi} will be in survey mode at all times, and that triggered rotations to view GRBs will not significantly raise the number of high-energy photons gathered.

The observations of GRBs by IACTs such as MAGIC are highly sensitive to the capabilities of the instrument.  While these telescopes have much larger effective collection areas than space-based instruments such as the LAT, other constraints such as the energy threshold, duty cycle, and time to respond to an alert must be taken into account in the analysis.  The much larger effective area of these telescopes compared to {\it Fermi} is compensated by the relatively small probability that any single event will be observable.  The effective area of IACTs is a strong function of energy in the sub--TeV regime, as the combined effects of the trigger efficiency and the reconstruction efficiency of low-energy showers in the analysis leads to a sharp decrease in effective area near threshold.  The potential of detecting GRBs is strongly affected by the low energy capabilities of the instrument, due to the rapidly increasing opacity of the universe to gamma rays above a couple hundred GeV for all but the closest bursts.  For MAGIC, we have used the `after cuts' form for the effective area as a function of energy from \citet{albert08a}, which is defined for angles near zenith.   For observations away from zenith, the telescope loses sensitivity at low energies due to the increasing amount of atmosphere through which the particle shower is being observed.  To estimate the change in the effective area function as a function of angle from zenith $\theta$, we use the following simple approximation:

\begin{equation}
A_{eff}(E,\theta)=A_{eff}(E-E',\; \theta=0),
\label{eq:magthresh}
\end{equation}
where
\begin{equation}
E'=50 \mbox{ GeV} \cdot( (\cos \theta)^{-2.5} - 1).
\end{equation}
\noindent This expression is based on a fit to the energy threshold results at lower angles from zenith ($\theta \lesssim 40$ deg) presented in \citet{firpoPHDT}.  Additionally, we implement in our analysis an absolute cut at energy $E_{min}$, below which the sensitivity is zero and no gamma rays will be observed.  We will perform our analysis for values of 50 and 100 GeV in $E_{min}$, to allow for uncertainty in low-energy sensitivity and data reconstruction.   Note that this parameter is distinct from the instrument energy threshold in that it is an absolute cut and is not dependent on the spectral properties of the source.  Recent estimates of the instrument sensitivity for stereo observations \citep{colin09} suggest that our more optimistic cut of 50 GeV may be reasonable for future observations.

A realistic estimate of the instrument duty cycle is critical for our analysis.  As outlined in \citet{bastieri05}, there are several requirements for the operation of MAGIC, including distance of the sun from zenith ($>108$ deg), a minimum angular distance of the moon from the observation field ($>30$ deg), and humidity and wind requirements.  For the duty cycle of the instrument, we use the standard value of 10 per cent, and note that this is supported by the fraction of GRBs (9.2 per cent) which were observed during follow-up in 42 months of observations \citep{garczarczyk08}, and is similar to reports from observations with other IACTs.  

A major challenge for ground-based attempts to detect GRBs has been the highly transitory nature of the emission.  Responding to an event requires the minimization of the individual components that contribute to the total delay time, including the time for the detecting satellite to confirm a burst in progress, the time to transmit this information, and then the time for the ground-based telescope to slew to the coordinates and begin taking data.  The first quantity depends on the satellite response time and strength of the burst.  Typical numbers for the {\it Swift} BAT are $<15$ sec to confirm and transmit a coordinate with a precision of a few arcminutes.  Communications of these coordinates are received by the IACTs in real time ($\sim 2$ sec, \citealp{bastieri05}) over the GRB Coordinate Network (GCN)\footnote{http://gcn.gsfc.nasa.gov/}.  However, the final step of repositioning a telescope the size of MAGIC on the time scale required to view prompt GRB emission presents a major engineering challenge.  Ensuring personnel safety is also a major practical concern in minimizing response time, which requires that the telescope be able to reposition without warning at any time during operation. 

\begin{figure}
\psfig{file=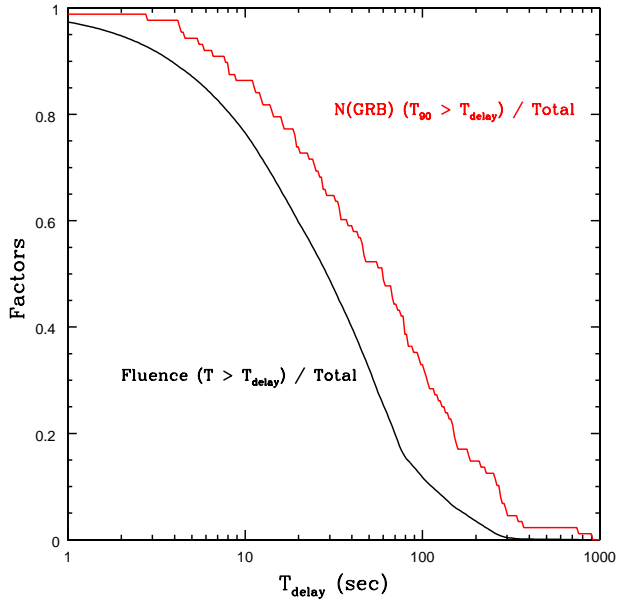,width=\columnwidth}
\caption{The fluence factors arising due to the slew time delay for ground-based telescopes.  The solid black curve shows the percentage of prompt fluence in the total sample which is seen after a time $T_{delay}$, if the GeV flux is proportional to that in the BAT energy range and is constant across $T_{90}$.  The red curve shows the fraction of GRBs for which $T_{90}$ is longer than $T_{delay}$.  This plot shows the percentage of high energy emission that will be missed by a telescope with a given delay time (due to slewing and other factors) in our model, after averaging over all {\it Swift} bursts.}
\label{fig:slewfact}
\end{figure}

For our analysis, we assume a delay time $T_{delay}$ in prompt observations, which incorporates all three times discussed above.  We assume a typical report time of 15 seconds for the burst alert to reach the instrument, and, in the case of MAGIC, a 30 sec slew time to move to the target and begin observations.  The total of 45 seconds is about equal to the lowest time, 43 sec, reported in \citet{garczarczyk08} for all of MAGIC GRB responses to date, and is therefore optimistic.  To compute the flux in the prompt phase of a GRB that can be seen by ground-based observations after this delay time we use the $T_{90}$ variable in BAT flux reported for our sample of bursts, which is the time elapsed between the arrival of the first 5 and first 95 percent of the total GRB fluence in the BAT band.  The fluence of the burst is modified by a factor 
\begin{equation}
F=F_{BAT} \cdot \mbox{MAX} \left[ \frac{T_{90}-T_{delay}}{T_{90}}\mbox{, }0 \right]
\end{equation}
that is, we take the prompt phase emission profile to be approximately flat over this timescale, and reduce the fluence by the proportion of the prompt phase that was missed.  Delays due to IACT response time do not affect the observation of afterglows.  We also do not account for the fact that on a timescale of hours the rotation of the Earth can bring afterglows into and out of the viewing region of the telescope.  In Figure \ref{fig:slewfact}, we show how the time delay in IACT observations affects observed flux in our model.  It is worth pointing out that we do not find $T_{90}$ to be correlated with either fluence or redshift in our sample; its correlation coefficients with these variables are -0.017 and -0.014, respectively.

\section{Results}
\label{results}
Based on the model for GRB emission and instrument performance developed in the previous section, we predict the number of high-energy gamma rays from GRBs that will be seen by {\it Fermi} and MAGIC, using the data set of {\it Swift} BAT GRBs with confirmed redshifts and the emission model discussed in the previous section.  In Table \ref{tab:params}, we review the parameters for the emission model and instrument properties that we are using in this section.
\begin{table}
\caption{Some of the parameters we use for calculations of this section.  See Sections \ref{grbemiss} and \ref{instprops} for more details.}
 \label{tab:params}
 \begin{tabular}{|l|c|l|}
\hline
$F_{\mbox{\tiny{BATSE}}}/F_{\mbox{\tiny{BAT}}}$ & 4.6 & BAT to BATSE fluence ratio \\
\hline
$\rho_{pr}$ & 0.1 & Fluence ratio for prompt phase \\
\hline
$\beta_{pr}$ & -1.95 & Prompt phase spectral index \\
\hline
$\rho_{ag}$ & 0.01 & Fluence ratio for afterglow phase \\
\hline
$\beta_{ag}$ & -1.5 & Afterglow spectral index \\
\hline
$T_{delay}$ & 45 sec & Delay time for MAGIC observations \\
\hline
\end{tabular}
\medskip
\end{table}

\subsection{Predicted Fluences for {\it Fermi}}
\label{fermipred}

With its wide field of view, Fermi is expected to view several high-redshift GRBs per year.  We therefore present results for this section in terms of broad bins in redshift, so as to have a reasonable statistical sample in each bin.  In Table \ref{tab:swiftsamp} we show the redshift bins, number of bursts seen in 54 months operations, and total fluence. 

\begin{table}
\caption{The redshift bins we use in our analysis and the numbers of GRBs and total fluence in each bin for the sample of {\it Swift} bursts.  The fluences shown are in the BAT energy range, 15--150 keV.  These data are over 54 months of observations, from January 2005 to June 2009.}
 \label{tab:swiftsamp}
 \begin{tabular}{|ccc|}
\hline
{\bf Redshift Bin} & {\bf N(GRB)} & {\bf Fluence ($10^{-7}$ erg cm$^{-2}$}) \\
\hline
$1<z<2$ & 29 & 1595.9 \\
\hline 
$2<z<3$ & 33 & 864.9 \\
\hline
$3<z<4$ & 21 & 665.1 \\
\hline
$4<z<6$ & 10 & 228.3 \\
\hline
\end{tabular}
\medskip
\end{table}

After calculating the high energy gamma-ray fluence for each burst in the sample, the flux is attenuated using optical depths calculated from the evolving background spectral energy distribution; see \citet{madau&phinney96} or \citet{gilmoreUV} for a review of these calculations.  In each redshift bin, we use attenuation factors averaged in redshift,
\begin{equation}
\frac{dF_{obs}}{dE}= \frac{dF}{dE} \: \int_{z_1}^{z_2} \frac{e^{-\tau(E,z)}}{z_2-z_1} \: dz,
\end{equation}
where $dF/dE$ is the high energy spectrum from Equation \ref{eq:heflux}, and $\tau(E,z)$ is the EBL optical depth as a function of observed gamma-ray energy and source redshift.  High-energy gamma rays are therefore assumed to originate from sources evenly distributed across the redshift bin.

\begin{figure*}
\psfig{file=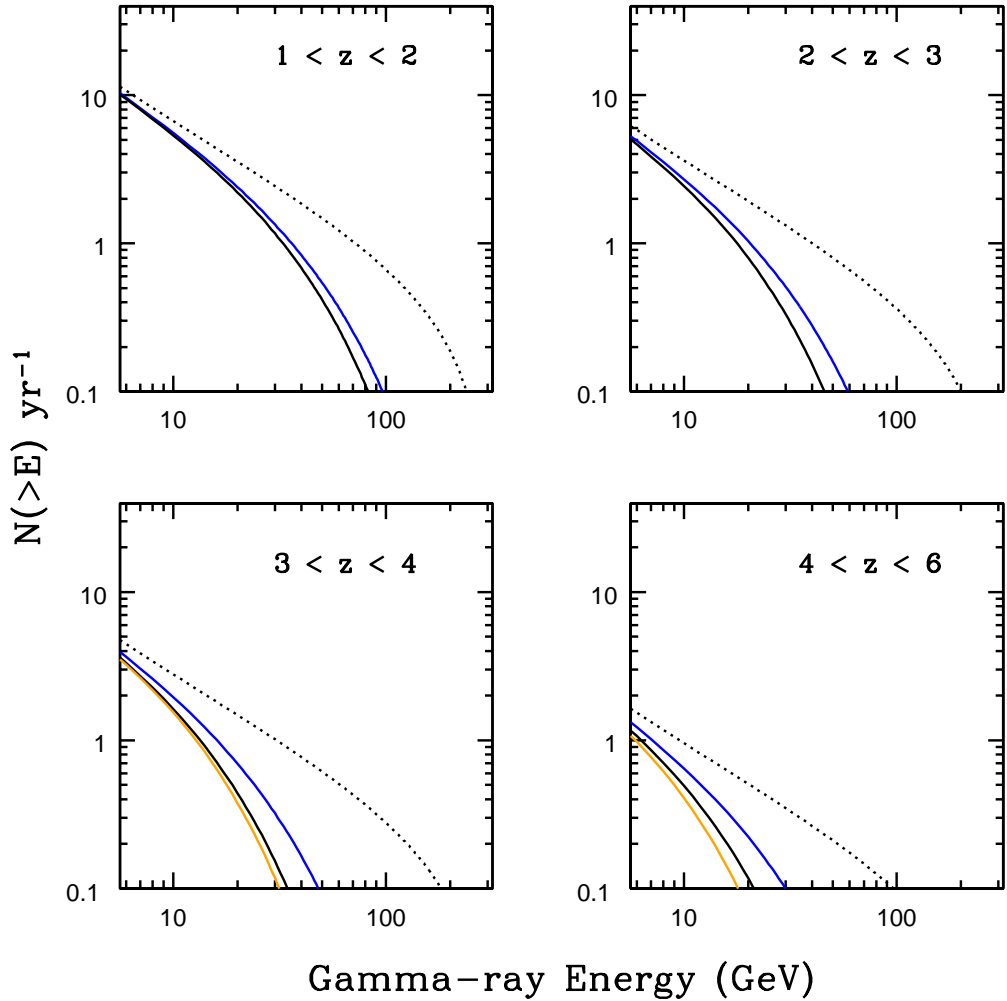,width=14cm}
\caption{The mean number of integrated photons above a given energy visible to {\it Fermi} per year, up to the maximum LAT energy of 300 GeV.  The dotted line shows the unabsorbed rate, while the solid blue and black (upper and lower, respectively) lines show results with attenuation due to the low and fiducial EBL models of \citet{gilmoreUV}.  The solid orange line (lowest line in the bottom panels) shows the integrated counts using the high model, which leads to greater opacity than the fiducial model above redshift 3.}
\label{fig:photint_fermi}
\end{figure*} 

\begin{figure*}
\psfig{file=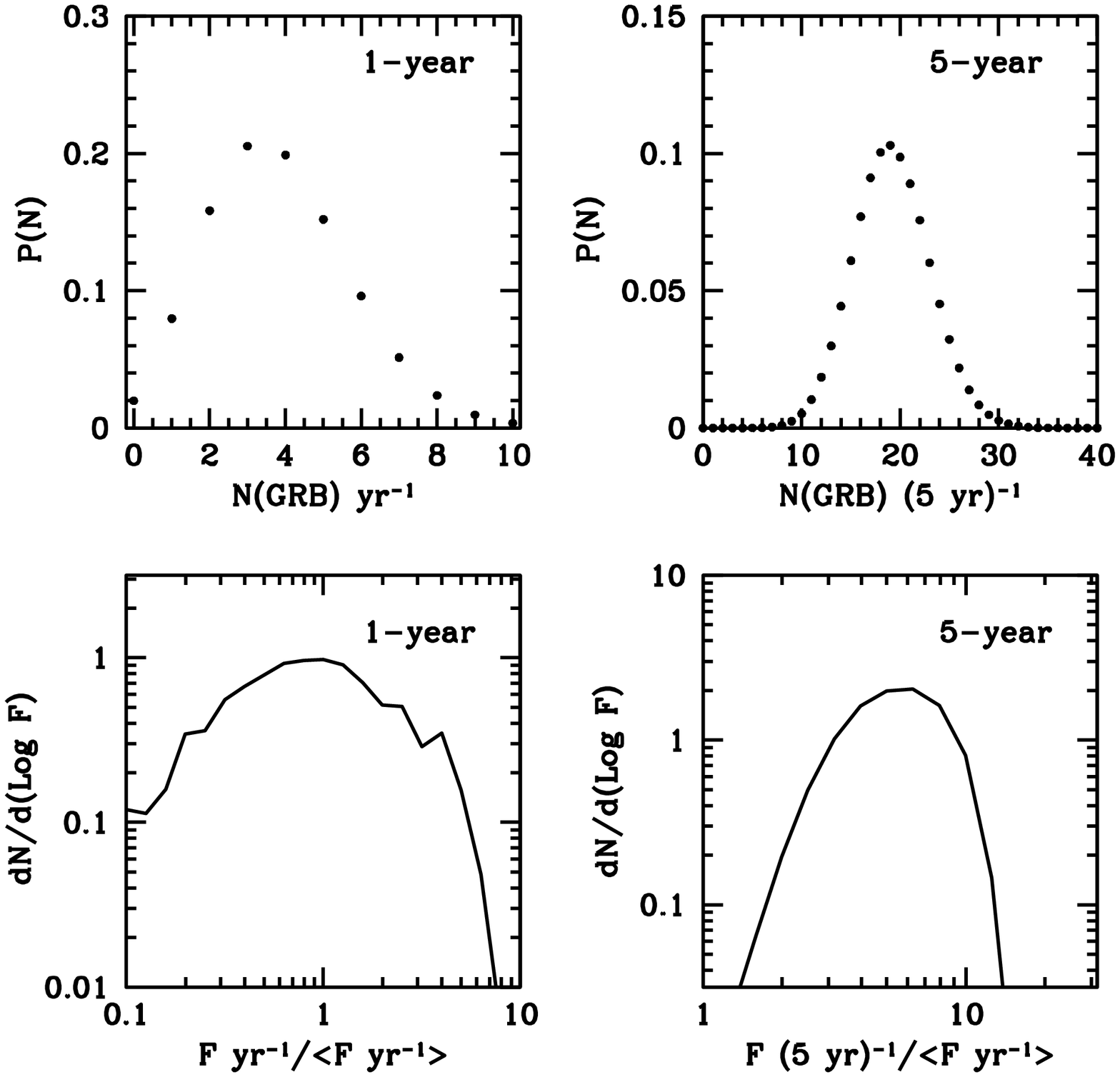,width=14cm}
\caption{The variation in yearly predictions for observations with the {\it Fermi} LAT of GeV gamma rays from GRBs with determined redshifts.  The upper-left plot shows the annual probability of a given number of GRBs occurring in view of the LAT (taken here to be angles $\leq 50$ deg from boresight) with redshifts between 1 and 6.  This does not take into account the expected number of GeV photons from each GRB, which may be less than one.  The lower-left plot is the yearly distribution of stacked fluences, normalized to the mean yearly predictions in Figure \ref{fig:photint_fermi}.  On the right-hand side we show the same quantities computed for a 5--year period. }
\label{fig:fourfermi}
\end{figure*} 

Figure \ref{fig:photint_fermi} presents the main results of this section; the mean number of photons above a given energy available without absorption by the EBL, and after absorption by our background models.  These predictions are made by combining our models for high-energy GRB emission for the {\it Swift} population and {\it Fermi} instrument properties, making adjustments for the field of view and effective areas for the two detectors.  Our results are divided into the four redshift bins, and show the averaged total amount of fluence from sources at these redshifts per year.

In Figure \ref{fig:fourfermi} we show how the mean fluence predictions can be expected to vary from year to year, based on the number of bursts in the {\it Swift} sample and the variance in fluence in this population.  This figure has been created using a simulation of LAT observations, and assuming randomly occurring bursts with the flux and redshift distribution of the {\it Swift} population.   The upper two distributions show the number of high-redshift ($1<z<6$) bursts falling in the field of view of the detector over a period of 1 (upper-left) and 5 (upper-right) years; this follows a Poisson distribution.  As before, we do not account for the possibility of the spacecraft autonomously slewing to view events with the LAT after being triggered by the GBM or another experiment.  The lower panels show how the stacked fluence collected can be expected to vary from the predictions in the previous plots.

\subsection{Predicted Fluences for MAGIC}
\label{magicpred}

\begin{figure}
\psfig{file=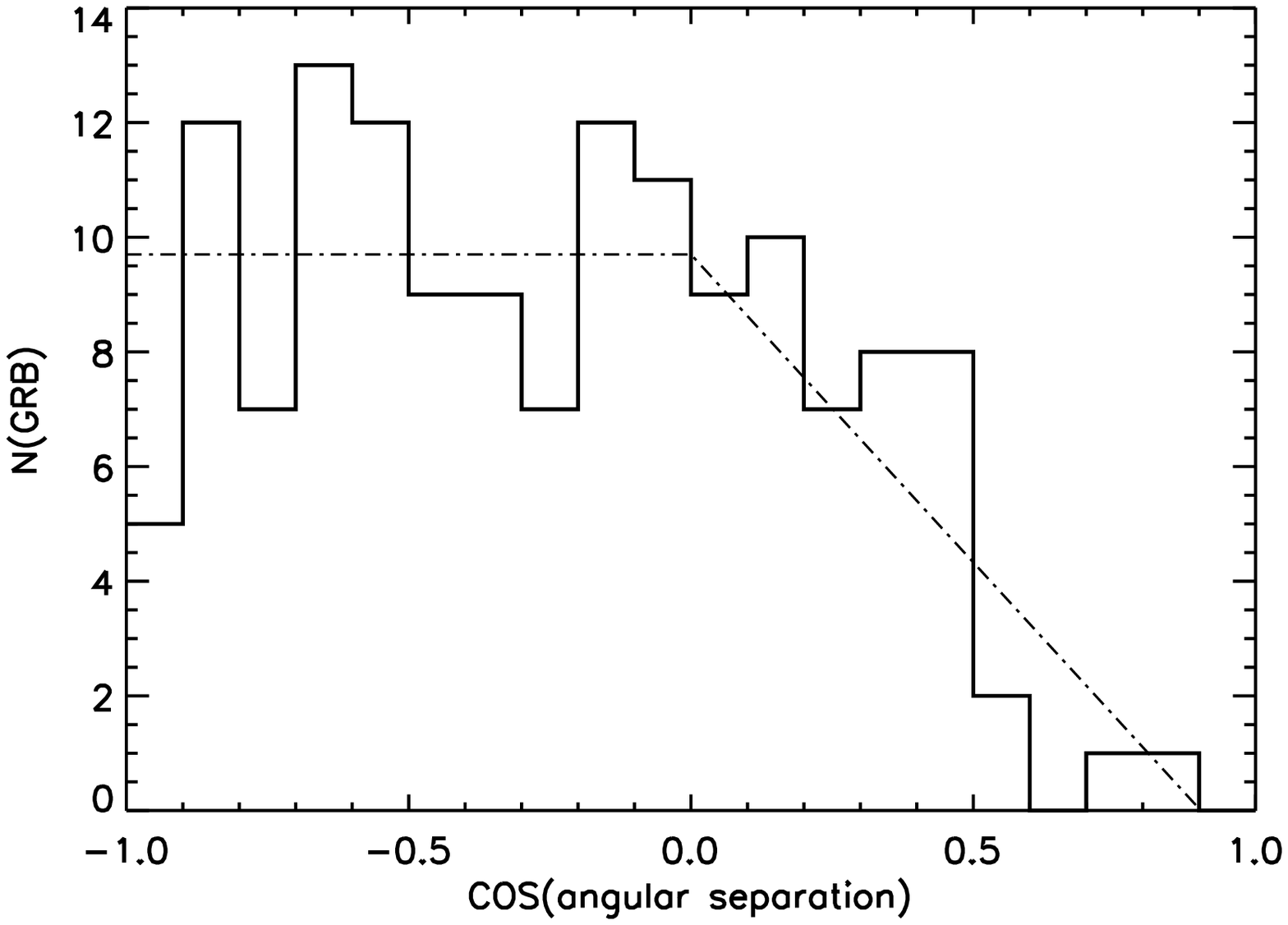,width=\columnwidth}
\caption{The distribution of the angular separation $\theta_{sep}$ between {\it Swift} GRBs with redshifts and the position of the sun.  This is plotted as $\cos(\theta_{sep})$, which would yield a flat distribution if the GRBs appeared independently of solar position.  The dot-dashed line is the fit used in determining the enhancement factor for IACT observations due to the solar anti-bias in burst alerts; see text for details. }
\label{fig:solardist}
\end{figure}

\begin{table*}
\caption{The annual probabilities of a given number of GRBs falling within a given angle from zenith, assuming a duty cycle of 10 per cent.  Under `afterglows', we include all bursts, and for `prompt' only those with $T_{90} > T_{delay}$.  Note that this calculation is for the number of GRBs falling in view of the instrument, and does not predict the number of expected photons, which may be less than one.  In Figs.~\ref{fig:magicgrb50} through \ref{fig:magicgrb_noag100} we show the distribution function for the number of photons received from a single burst falling in view of the instrument, using our emission model and taking into account EBL attenuation and instrument effective area.  These factors also take into account the antisolar bias found in the distribution of {\it Swift} GRBs. }
 \label{tab:magicgrbyr}
 \begin{tabular}{|c|ccccccc|}

\hline
 &  &\hspace{0cm} Prompt ($T_{delay}=45 s$) & & \hspace{0.6cm} & & \hspace{0cm} Afterglow ($T_{delay}=0$) & \\
\hline
 N(GRB) yr$^{-1}$ & 20 deg & 30 deg & 40 deg & \hspace{0.6cm} & 20 deg & 30 deg & 40 deg \\
\hline
0  & 0.95 &  0.90 & 0.82 & \hspace{0.6cm} & 0.92 & 0.82 & 0.71  \\
\hline
1 & 0.048 & 0.099 & 0.16 & \hspace{0.6cm} & 0.079 & 0.16 & 0.24  \\
\hline
$\geq 2$ & 0.0012 & 0.0057 & 0.016 & \hspace{0.6cm} & 0.0036 & 0.016 & 0.046  \\
\hline
\end{tabular}
\medskip
\end{table*}
% from calculation:
% 0.91695597      0.82441461      0.71378565      0.95129669      0.89501101      0.82390928
%     0.079487133      0.15919589      0.24069390     0.047502488     0.099284269      0.15960355
%    0.0035568923     0.016389504     0.045520455    0.0012008250    0.0057047233     0.016487166

\begin{figure}
\psfig{file=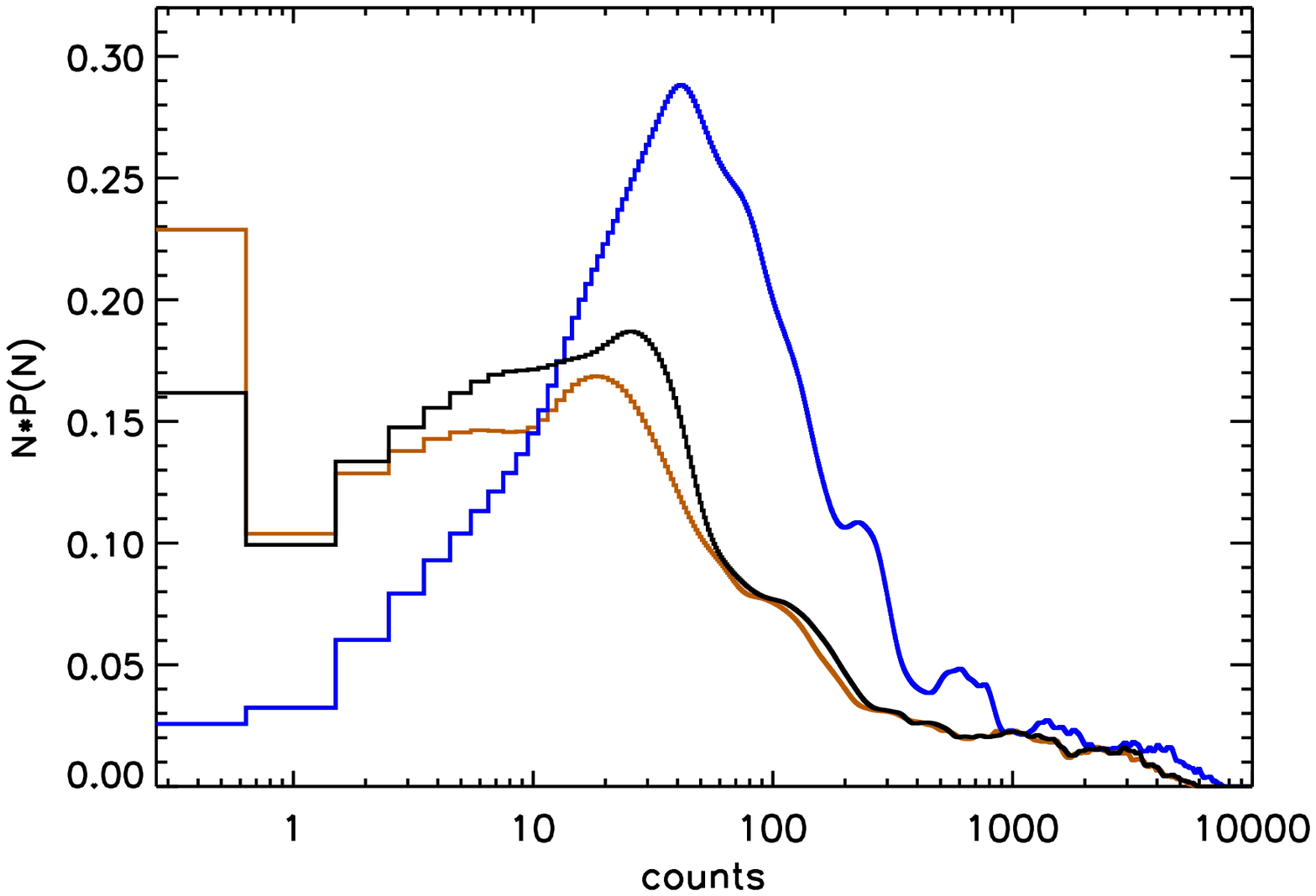,width=\columnwidth}
\caption{Predictions for the number of gamma-rays visible to MAGIC from a high-redshift ($z>1$) GRB; see Table \ref{tab:magicgrbyr} for the rates at which these events are expected to occur in view of the telescope.  Here we use parameters $T_{delay}$=45 sec and $\theta_{max}=40$ deg, and assume minimum energy of 50 GeV in telescope sensitivity.  Blue, black and orange lines show the results after attenuation by the low, fiducial, and high EBL models, respectively.  The leftmost bin is the probability of zero photons being received.}
\label{fig:magicgrb50}
\end{figure} 

For MAGIC, we begin by considering the year-to-year probability that a given number of high-redshift GRBs will occur in a region of sky where they can observed with a low energy threshold.  By multiplying the duty cycle of the instrument (10 per cent) with the sky coverage (11.7 per cent for $\theta_{max}= 40$ deg), we find that only $\sim$1 per cent of bursts can be observed with a reasonably low energy threshold.  Our sample consists of 145 bursts seen over 54 months, 96 of which are at $z>1$, and therefore the expected number of bursts per year is somewhat less than 1.  Therefore, we begin by predicting the probability that any bursts will be visible.  Once we understand this probability, we will look at the photon statistics for a single burst.  

The sky coverage of the telescope increases approximately as the square of the maximum allowed angle from zenith.  However, we find that the number of photons predicted from distant ($z>1$) bursts does not increase significantly beyond an angle of about 40 degrees, as the energy threshold of the instrument rises above the energies at which the universe is transparent to gamma-rays.  GRBs which occur far from zenith are therefore shrouded from view by EBL attenuation unless they are at low redshift.  The number of bursts per year will also depend on whether prompt or afterglow phases are being considered, as the $T_{90}$ duration of some GRBs will be less than $T_{delay}$, preventing most or all of their prompt emission being viewed.  In Table \ref{tab:magicgrbyr} we present probabilities for the number of GRBs visible to MAGIC per year, within a given maximum angle from zenith. 

Another factor that must be taken into account for this analysis is the anti-solar bias in the distribution of {\it Swift} GRBs.    As the determination of redshifts from afterglow observation is hampered by glare from the sun, {\it Swift} preferentially finds GRB events in the anti-solar direction.  This works to the advantage of IACTs, which can only operate at night, by effectively increasing the duty cycle with respect to {\it Swift}-triggered bursts.  As low-threshold IACT observations are limited almost exclusively to solar angles of $>$90$^{\circ}$, we can estimate the effect of this bias factor on the duty cycle by taking the GRB distribution to be flat for separation angles $>$90$^{\circ}$ and modeling a cutoff in GRB detection rates at smaller angular separations.  In Fig. \ref{fig:solardist} we show the distribution of bursts on the sky with respect to the position of the sun.  Using the simple fit shown, we find an enhancement factor of $\approx 1.38$ for IACTs relative to an isotropic distribution. 

We find that the probability of a high redshift burst falling within the field of view of an IACT in any particular year is small, $\sim$30 per cent for $\theta_{max}=40$ deg.  At this angle, the energy threshold is approximately double that at zenith, and rises rapidly for larger angles \citep{firpoPHDT}.  It is not realistic to expect to see GRBs at $z>1$ at larger zenith angles for either of our EBL models due to attenuation, although low redshift bursts could be visible if photon emission takes place at high enough energies.  In years in which a nonzero number of bursts is in view of the telescope, the predicted flux is expected to vary highly due to the large range of fluences seen in the {\it Swift} sample (y-axis in Figure \ref{fig:batfluxdist}).  It is therefore not particularly useful to describe an `average' year, as we did in the previous section when discussing {\it Fermi}, as the median number of observed gamma-rays from GRBs in this case is zero. 

\begin{figure}
\psfig{file=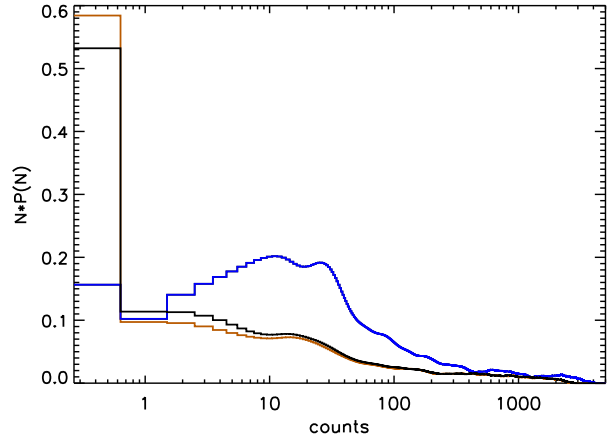,width=\columnwidth}
\caption{As in the previous figure, but with $E_{min}=100$ GeV}
\label{fig:magicgrb100}
\end{figure}

\begin{figure}
\psfig{file=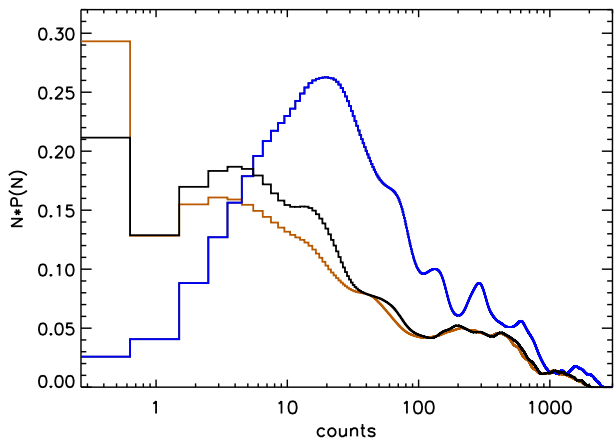,width=\columnwidth}
\caption{Result for observations with an energy cut at $E_{min}=50$ GeV, and here including only photons from the prompt phase of emission.  Line types are as in Figure \ref{fig:magicgrb50}.}
\label{fig:magicgrb_noag50}
\end{figure}

\begin{figure}
\psfig{file=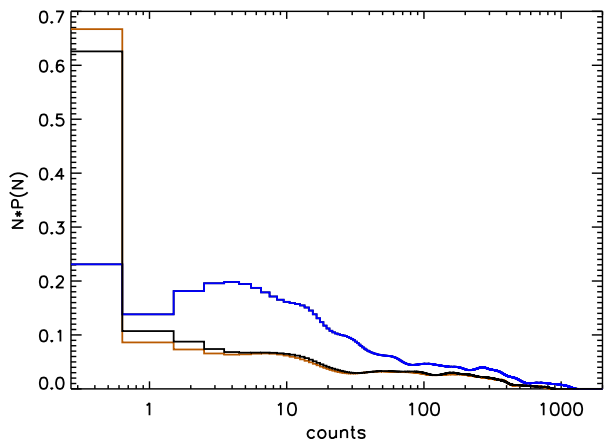,width=\columnwidth}
\caption{Result for observations with $E_{min}=100$ GeV, and including only photons from the prompt phase of emission.  Line types are as in Figure \ref{fig:magicgrb50}.}
\label{fig:magicgrb_noag100}
\end{figure}

In Figures \ref{fig:magicgrb50} and \ref{fig:magicgrb100} we show as a histogram the probability distribution for the number of detected photons expected from a single distant ($z\geq 1$) GRB within $\theta_{max}$.  This calculation has been done by repeatedly sampling our catalogue of {\it Swift} bursts, randomizing the sky position of the GRB within the disk of radius $\theta_{max}$ each time. The MAGIC effective area as a function of energy is then calculated using Equation \ref{eq:magthresh}.  We apply absolute cuts in energy of $E_{c} = 50$ and $E_{c}=100$ GeV to our analysis, respectively in these two figures.   The predicted high energy flux is attenuated using our three evolving background models.  Additionally, the redshift of each burst is randomized each time it is sampled by an amount $-0.2 \leq \Delta z \leq 0.2$.  Figures \ref{fig:magicgrb_noag50} and \ref{fig:magicgrb_noag100} shows the same plots including only the prompt phase photons.

\begin{figure*}
\psfig{file=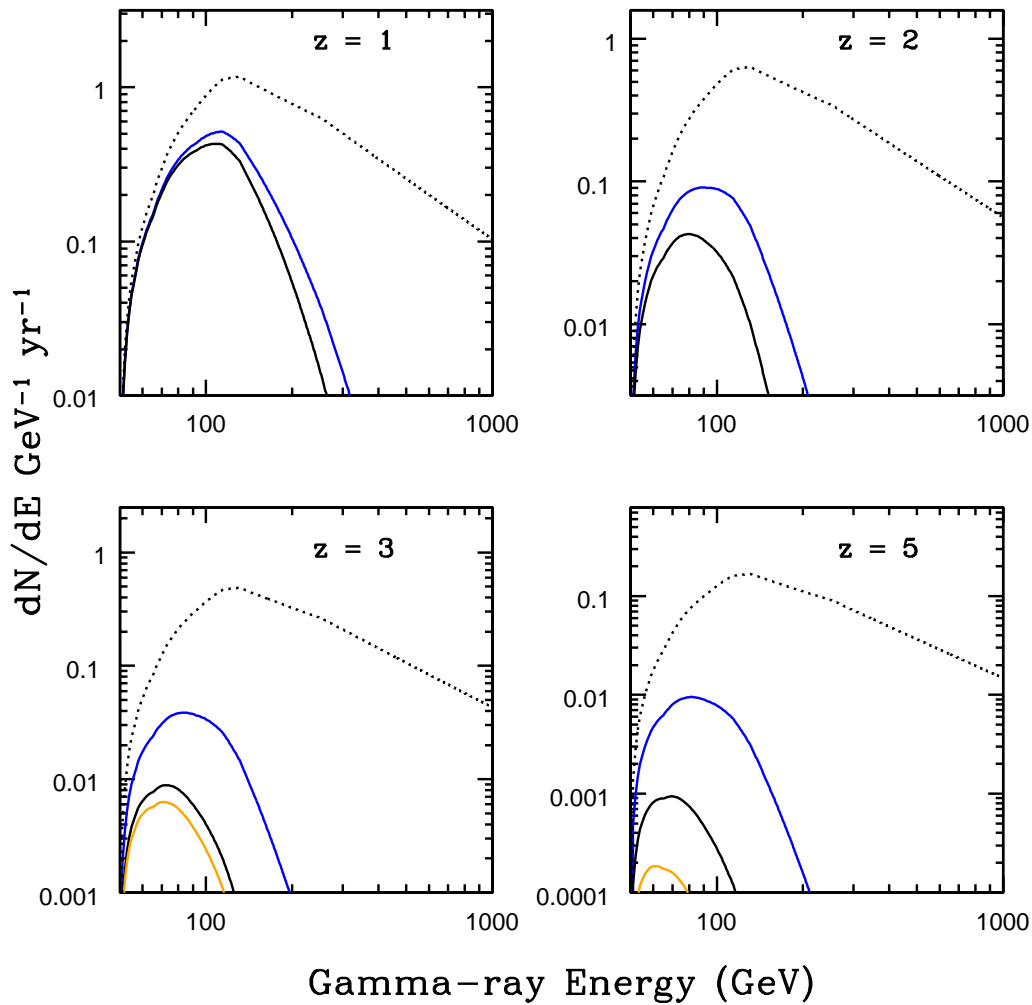,width=14cm}
\caption{Predictions for the spectrum observable by MAGIC from distant GRBs, at 4 different redshifts.  Here we assume observations at 20 degrees from zenith with an $E_{min}=50$ GeV.  The dotted black line is the unattenuated spectrum after including instrument effective area, and the solid orange, black, and blue lines (lower to upper) show the flux after attenuation by the high, fiducial, and low EBL models.  The fluence is normalized to the average per year, as was done for our {\it Fermi} results.  However, as we have seen, the year-to-year prediction is highly variable, so the vertical scale of the plot should be taken as arbitrary, and Figs.~\ref{fig:magicgrb50} through \ref{fig:magicgrb_noag100} used as a gauge of expected number of counts.}
\label{fig:photdiff_mag}
\end{figure*}

Figure \ref{fig:photdiff_mag} shows the expected $dN/dE$ spectrum of photons arriving from GRBs at different redshifts.  This plot takes into account the assumed gamma-ray spectrum, the instrument effective area, and EBL attenuation for four different redshifts.  For MAGIC, we find that photons are expected to be seen in a fairly narrow energy range peaking at $\sim$ 100 GeV.  The number of photons expected near the minimum energy of the instrument are suppressed due to the small effective area at these energies.  At higher energies, the spectrum declines rapidly due to EBL attenuation.  As none of these spectral factors depend on absolute fluence in our model, these results are valid regardless of the luminosity of the GRB.  We have normalized the spectra to the average yearly flux for convenience, however this normalization in itself has little meaning due to the large amount of variance in our predictions.

\section{Discussion}
\label{disc}
We have attempted to make predictions for the number of high-energy gamma rays that can be seen by current generation of telescopes targeting the GeV energy range.  Although we have strived to design a simple and straightforward model, there are a large number of uncertainties in predicting the high-energy emission of GRBs.  Over the next few years, observations in the GeV energy band will be able to constrain many of the assumptions that we have used here.  

Our findings suggest that the {\it Fermi} LAT will typically observe at least 3 to 4 bursts per year with redshifts that are determined to be greater than 1.  Over the lifetime of the mission ($>$5 yr), this means that multiple events will likely be seen in each of the redshift bins.  While not all these GRBs will necessarily have detectable GeV emission, the stacked results could yield enough photons above 10 GeV to place constraints on the EBL, and possibly differentiate between the two models we have presented here.  It is difficult to make generalizations about yearly predictions, due to the large amount of variance in both number of bursts viewed and fluence per individual burst, which vary over orders of magnitude.  Our predictions become more stable over the 5-year instrument lifetime (right-hand panels of Fig. \ref{fig:fourfermi}); the probability of our results varying by more than a factor of $\sim 2$ from predictions on this timescale is small, at least when totals from all redshift bins are considered (this is likely smaller than the expected variation in $\rho$ from Section \ref{grbemiss}).  We predict, on average, several photons per year above 10 GeV for $1 \leq z \leq 2$ and two or three from $2 \leq z \leq 3$.  This is comparable to the more optimistic predictions of \citet{le&dermer09}, who considered all long-duration bursts over the full sky with the same high energy spectrum and MeV--GeV fluence ratios.    We have not included autonomous repoints of the instrument in response to a GBM trigger in our model.  If repoints are performed frequently and prove an effective way to view GeV emission from bursts, then this could effectively increase the LAT field of view to $1/2$ the sky, a factor of 2.5 higher than we consider.

For an IACT like MAGIC, the annual likelihood of viewing one or more distant GRBs within 40 degrees of zenith is less than 30 per cent.    We have chosen this cutoff due to the rapidly growing energy threshold at higher angles, and the fact that the EBL absorption attenuates most emission at these energies for sources at $z>1$ in all of our EBL models.  As shown in Figure \ref{fig:photdiff_mag}, the majority of photons from these bursts can be expected to arrive at $\sim 100$ GeV, or lower for high redshift sources.  This means that it is the capabilities of the detector near to the lower end of its energy range that are most important to viewing distant GRBs, not the energies at which the instrument necessarily has the most sensitivity.  As can be seen in Figs. \ref{fig:magicgrb100} and \ref{fig:magicgrb_noag100}, increasing the assumed minimum energy of the instrument to 100 GeV greatly decreases sensitivity to GRBs, especially when combined with a higher realization of the EBL.  Sensitivity below 100 GeV is therefore critical to detecting high redshift GRBs if the UV-optical background is as high or higher than found in our fiducial model.

In this analysis we have restricted ourselves to GRBs at $z \geq 1$.  For GRBs at lower redshifts, more photons may be observable in the 100 GeV to 1 TeV decade, but these bursts only represent a minority of those for which we have redshifts.  Another reason for focusing on the 10 to $\sim$100 GeV decade is that we are relying on EGRET observations at maximum energies of $\sim 10$ GeV to model our high-energy emission, and we expect this model to become increasingly uncertain at higher energies.  Our main results from this section show the potential of receiving a large number of photon counts from a single GRB.  Even a relatively small number of gamma rays could be very useful in constraining the EBL through its effects on the spectrum and total power of the VHE emission.  Our results suggest that there is a large degree of chance involved in seeing GRBs from the ground, but there could be a large payoff for our knowledge of cosmology from even one success.

For both MAGIC and {\it Fermi}, we have based our calculation on the population of bursts seen by {\it Swift} for which redshifts were eventually determined.  However, {\it Fermi} is a capable finder of transients, covering approximately 1/5$^{th}$ of the sky with the LAT and possibly more after GBM-triggered repointings are taken into account.  Above 10 GeV, the LAT has an angular resolution of $\leq 0.1$ degrees, allowing strong bursts to be targeted for multiwavelength observations.  It is therefore possible that {\it Fermi} will enable the determination of many GRB redshifts without simultaneous {\it Swift} detections, and the number of high-redshift bursts we have predicted for the next few years could be an underestimate.  This could improve the prospects for detecting GeV photons from these sources with both {\it Fermi} and IACTs. 

\subsection{Simulated Results for GRB 080916C}
\label{080916C}
One exciting implication in our findings is the potential payoff from a single bright GRB.  For MAGIC, we find that while the probability of seeing photons from any single event is quite small, the reward for catching a burst of intermediate to high fluence could be hundreds or thousands of photons observed within a narrow energy range.  The detailed spectrum from such an event could be invaluable for constraining the UV background and high-redshift galaxy formation.  Just as a demonstration of how many photons one event could provide, we consider recent GRB 080916C which was observed in its full prompt phase by the {\it Fermi} LAT, as well as GBM.  {\it Swift} did not observe this event until nearly a day afterwards, and it was not included in the calculations of the previous section.  This burst, which occurred on September 16, 2008, is among the brightest GRBs ever seen, and with a redshift of $4.35\pm 0.15$ it is the most energetic burst currently known \citep{greiner09}.  As described in \citet{abdo09}, it was seen by the LAT at an angle of 48 degrees from boresight following a trigger from the GBM.  A total of 145 gamma rays above 100 MeV and 14 gamma rays above 1 GeV were reported.   The last of those 14 gamma rays arrived approximately 46 seconds after the initial trigger.  The highest energy gamma ray was measured to be $13.22^{+0.77}_{-1.54}$ GeV, and occurred 16.54 seconds after the trigger.

A useful test of our emission model is to ask if we arrive at similar results for the number of LAT-observed photons above 1 GeV, using the GRB 080916C GBM fluence of $1.1 \times 10^{-4}$ erg cm$^{-2}$ as the basis for our calculation.  Following the same analysis as in Section \ref{fermipred} for this single GRB, we predict 24 and 23 photons above 1 GeV respectively for our low and high EBL models, compared to 14 observed by the LAT.  For photons above 10 GeV, we predict 1.9 and 1.4 photons for the two models, compared to one seen.  The number of predicted cumulative counts falls below 1 for energies of 15 and 11 GeV.  Thus, while our emission model overpredicts the number of photons around 1 GeV, it does a reasonably good job of predicting the highest energy photon seen by the LAT for this event, though it is difficult to draw conclusions with such small statistics.  The role of attenuation by the EBL in this energy range is minimal.  Our low and high models give optical depths of 0.06 and 0.22 for a photon of 13.2 GeV from redshift 4.35, corresponding to attenuation factors ($e^{-\tau}$) 0.94 and 0.80. 

Next, we consider a hypothetical observation of this GRB by MAGIC.  As the last photon from GRB 080916C above 1 GeV arrived at about 46 seconds after the initial trigger, it is unlikely that MAGIC or other IACTs could have seen much of the prompt emission, and the counts we calculate here are entirely from a hypothesized afterglow component.  For an IACT, observations of this burst would have been heavily dependent on the assumed EBL model, due to its high redshift.  We calculate that with a 50 GeV energy cut at zenith MAGIC would have seen 350 gamma rays for the EBL in our low model; in the fiducial and high models 58 and 19 gamma rays would be seen.  This assumes ideal viewing conditions, with the GRB occurring directly overhead, and in reality the chances of such an occurrence are exceedingly small.  At higher zenith angles, the number of observable photons declines rapidly due to the higher threshold and the EBL attenuation being a strong function of energy.  If the event were instead seen at 45 degrees from zenith the predicted gamma-rays counts for those same three EBL models would be 61, 2, and 0.24.  As we see, predictions can vary enormously for high-redshift GRBs depending on the background model.  The fiducial and high models create a dense background of UV photons due to earlier star formation, and 100 GeV gamma-rays from a source at $z \sim 4$ are attenuated by a factor $>100$.  Our low model has much less star formation at high redshift, and the optical depth to gamma rays is much lower, although the universe is still optically thick ($\tau > 1$) at this redshift for photons above 50 GeV.

As mentioned in \citet{abdo09}, the high energy emission measured by the LAT was delayed slightly compared to the GBM flux.  The highest energy photon, and two others that had energies above 6 GeV, did not arrive until more than 83 per cent of the prompt GMB fluence had been received (as seen in Table 1 of this reference).  While it is difficult to draw conclusions from one event, this may indicate that the VHE photons produced in the prompt phase may arrive later than the lower energy fluence which defines $T_{90}$, or possibly that the spectrum hardens with time and GeV photons tend to arrive later than lower energy emission.  We have assumed in our analysis that MeV and GeV prompt-phase flux are directly proportional in time with a constant spectrum.  If there is a delay or spectral hardening it could work to the advantage of ground-based instruments, allowing them more time to react to a GRB report than we have assumed here.  %

\subsection{Intrinsic Spectral Cutoffs}
\label{intcut}
One hurdle in detecting gamma-ray attenuation features could be the existence of a spectral cutoff due to either the Klein-Nishina cutoff or internal absorption of gamma rays.  As described in Equation \ref{eq:kncut}, the relevant energy scale is determined by the bulk Lorentz factor of the GRB ejecta and the typical electron relativistic Lorentz factor.  The analysis of GRB 080916C suggests a bulk factor of $\Gamma_{bulk} \geq 887 \pm 21$ during the time intervals when the highest energy gamma rays were emitted.  If the electron Lorentz factor was at least $\sim 10^3$, then emission that could be observed by MAGIC would not be affected.  However, as the most powerful GRB on record, parameters for GRB 080916C may not be representative of the total sample.  One potential danger is that the typical energy of the cutoff could exist at roughly the same GeV energies where we expect EBL attenuation features to be seen.  Not only could a sharp spectral cutoff be mistaken for attenuation by background radiation, but the factor of $(1+z)^{-1}$ from cosmology could mimic the redshift evolution of EBL attenuation.

\subsection{Future Experiments}
One reason we have restricted ourselves to current experiments in this discussion is that, as we have seen, the details of instrument capabilities can have a large impact on predictions, and our results are most meaningful when we can incorporate well-tested and verified instrument parameters into our model.  But as the GeV emission of GRBs and subsequent extinction are certainly not questions that are going to be decisively answered by the current generation of instruments, our discussion would not be complete without mentioning briefly a few important upcoming experiments.  The next phase of the H.E.S.S. array will feature a 600 m$^2$ mirror at the center of its current 4-telescope configuration; this central `T5' telescope will be the largest IACT yet built.  This upgrade is scheduled for completion later this year, and will lower the energy threshold down to $\sim$30 GeV at zenith angle 18 degrees \citep{becherini08}. Over the next decade, several ground-based experiments can provide more sensitivity to VHE photons from GRBs \citep{williams09}.  The Advanced Gamma-Ray Imaging System (AGIS) \citep{buckley08} and Cherenkov Telescope Array (CTA) \citep{martinez08} are two future concepts for IACT arrays that may be constructed during the next decade.  Both of these arrays, when fully constructed, would have much larger collection areas than any current experiment, and would likely have energy coverage over most of the 10 to 100 GeV decade.  The low energy threshold, $<20$ GeV, and greater sensitivity will enable detection of GRBs and measurement of attenuation by the EBL out to very high redshift, which could clarify the nature of high-redshift star formation.  Unfortunately, these telescopes will not be able to overcome the intrinsic difficulties of the Cherenkov technique, namely low duty cycle, loss of sensitivity away from zenith, and the need to be triggered for transient observations by another experiment.  Our results suggest that due to the stochastic nature of GRBs, persistence may ultimately be the key to detecting these events from the ground.

\section*{Acknowledgments}
R.G. and J.P. acknowledge support from a Fermi Guest Investigator Grant and NSF-AST-0607712, and J.P. and F.P. acknowledge support from a NASA ATP grant.  We thank Stefano Covino, Markus Garczarczyk, Markus Gaug, Nepomuk Otte, Stefano Profumo, and David A.~Williams for helpful discussion and comments.  We are particularly grateful to Nepomuk Otte for providing a great deal of assistance in understanding the MAGIC telescope, and to David A.~Williams for making us aware of the antisolar bias present in {\it Swift} observations, an important factor in our calculations related to ground-based instruments.

\bibliographystyle{apj}

\label{lastpage}
\end{document}